\documentclass{eas}
\usepackage{graphicx}
\begin{document}

\title{The mic-mac connection inside stars: \\
the interdependence of atomic diffusion and hydrodynamic instabilities.} 
\author{Vauclair, S.}\address{Universit\'e de Toulouse, UPS-OMP, IRAP, France and CNRS, IRAP, 14 avenue Edouard Belin, 31400, Toulouse, France }

\begin{abstract}
The interdependence of microscopic (atomic) and macroscopic (hydrodynamic) processes inside stars and their consequences for stellar structure and evolution were recognized by Jean-Paul Zahn several decades ago. He was a pioneer in that respect, discussing the importance of the macroscopic motions related to stellar rotation, in competition with the chemical stratification induced by gravitational settling and radiative accelerations. This has been much developed in recent years, in connection with the improvement of observational data, including asteroseismology. Morover, it has been recently discovered that the microscopic atomic diffusion processes can lead to macroscopic results which may infuence in a non negligible way the internal stellar structure, independently of the abundances observed at the surface. 
\end{abstract}
\maketitle

\section{Introduction}

Jean-Paul Zahn was a pioneer in stellar hydrodynamics and we had many discussions during several decades about the competition between microscopic atomic diffusion and various kinds of macroscopic mixing, including rotational mixing and internal waves. It appeared at that time that, when atomic diffusion is computed in stellar models without including any macroscopic mixing, it leads to abundance anomalies much larger than those currently observed in stellar atmospheres. The competition between microscopic and macroscopic processes seemed a necessity to account for the observations. A first discussion on that subject appeared in 1983, when Arthur Cox, myself and Jean-Paul Zahn were invited by Andr\'e Maeder to give reviews on stellar oscillations, atomic diffusion and mixing processes at the Saas Fee winter school. The book which appeared after that school has been used by many research students since that time (Cox, Vauclair \& Zahn \cite{cox83}).

More recently, it was realized that in some cases, the accumulation of heavy elements induced by atomic diffusion inside stars could lead to fingering (thermohaline) convection, due to the presence of inverse (unstable) mu-gradients. This special mixing is a direct consequence of the development of chemical gradients inside stars. It was not taken into account in the computations during all these years, simply because people did not think about it. It appears as an intrinsic abundance regulator, which decreases the abundance variations without suppressing them. The consequences of these combined processes are important for the internal stellar structure and the overall opacities. In the future, fingering convection must be treated together with rotational induced mixing for a better description of stellar interiors. At the present time, we still study them separately to analyse in detail their respective behaviors.

In the following, I recall the basics of atomic diffusion inside stars and I describe fingering convection and its importance for stellar structure. I then give some discussion about rotational induced mixing before concluding.

\section{ Atomic diffusion}

Atomic diffusion is a fundamental process related to the physics of stellar interiors. The relative motion of the various components of the stellar gas during the star's life can only be avoided by strong mixing, like in convective zones. In the radiative layers, this effect and its consequences cannot be neglected. The reasons why the stellar structure inevitably leads to selective diffusion of the various elements may be summarized as follows:

\begin{itemize}

\item {Stars are self-gravitating spheres. For this reason, they develop pressure, density, temperature gradients in their interiors,}

\item {Stars are made of multicomponent gases, with different atomic masses and atomic structure, }

\item {Because of their different physical characteristics, these gaseous components behave differently in the presence of the structural gradients.}

\end{itemize}

This effect is neglected in standard models and the equations are written as if the stellar gas was unique. The fundamental hydrostatic equilibrium equation is written by introducing an average atomic mass, the mean molecular weight $\mu$. Similarly the radiative transfer is computed inside the star as for a unique gas, which would have an average behavior with respect to the photons. For the sake of the computations, an average absorption coefficient is introduced in the corresponding equation: the Rosseland mean opacity.

In real stars:

\begin{itemize}

\item {Each gas behaves with its own molecular weight, while it feels the global pressure gradient}

\item {Each element absorbs photons in its own way, and feels the corresponding upwards momentum}

\item {Because of that, the various atoms move up or down with respect to one another, and all the acquired momentum is shared to the surroundings after a collision time scale.}

\item {Atomic diffusion corresponds to what happens during this collision time scale.}

\end{itemize}

Although recognized by the pioneers of the studies of stellar structure, the process of atomic diffusion in stars was forgotten during several decades. It was revived later on as the reason for the 
large variety of chemical peculiarities observed in main sequence and hot
horizontal branch stars (Michaud \cite{michaud70}, Michaud et al. \cite{michaud76}, Vauclair \& Vauclair \cite{vauclair82}). 
More recently, it was realized that the importance of this process is not restricted to the so-called ``peculiar stars"
but that it occurs in all kinds of stars, although moderated by
macroscopic motions and mixing. A spectacular confirmation of atomic diffusion inside the Sun was given by helioseismology
(Gough et al. \cite{gough96}, Richard et al. \cite{richard96}).

At the present time, many stellar evolution codes include gravitational and thermal diffusion
in their routines, but only very few include the computations of radiative
accelerations, which are mandatory to evaluate correct diffusion velocities. The computations 
of radiative accelerations are complicated indeed and time consuming. They involve
detailed investigations of all the atomic parameters of all the elements in all ionization
stages. For this reason, they are deeply related to the computations of stellar opacities.

Among the very few codes which include these computations, two of them have been directly compared by Th\'eado et al. \cite{theado09}, Th\'eado \& Vauclair \cite{theado12} and by Deal, Richard \& Vauclair \cite{deal16}. They are the
Montreal code, in which the computations of the radiative accelerations have been introduced in a detailed and precise
way, in connection with the OPAL atomic data (Iglesias \& Rogers \cite{iglesias96}), and the Toulouse code (TGEC) in
which radiative accelerations have been introduced by using the approximate method derived by Alecian \& LeBlanc\cite{alecian04} and LeBlanc \& Alecian \cite{leblanc04}. 
The Montreal code is more precise for radiative
accelerations than the TGEC one, but TGEC uses a less time consuming method
allowing for a better introduction of macroscopic motions.

The variations with depth of the radiative accelerations on specific elements lead
to their accumulation or depletion in various layers inside the stars, which may have strong
consequences on the stellar structure and evolution.
Typically, element accumulation occurs just above the place where they are the main contributors of the overall opacity. The reason is simple: the elements are main contributors to the opacity in the region where they absorb photons the most easily, due to their ionisation stage. In this case, their radiative acceleration is large and they are pushed up by the photon flux. Above that region, their ionisation stage changes and they absorb less photons. This creates in a natural way an accumulation of these elements in that special region (Fig. 1).

\begin{figure}
\includegraphics[width=12.5cm]{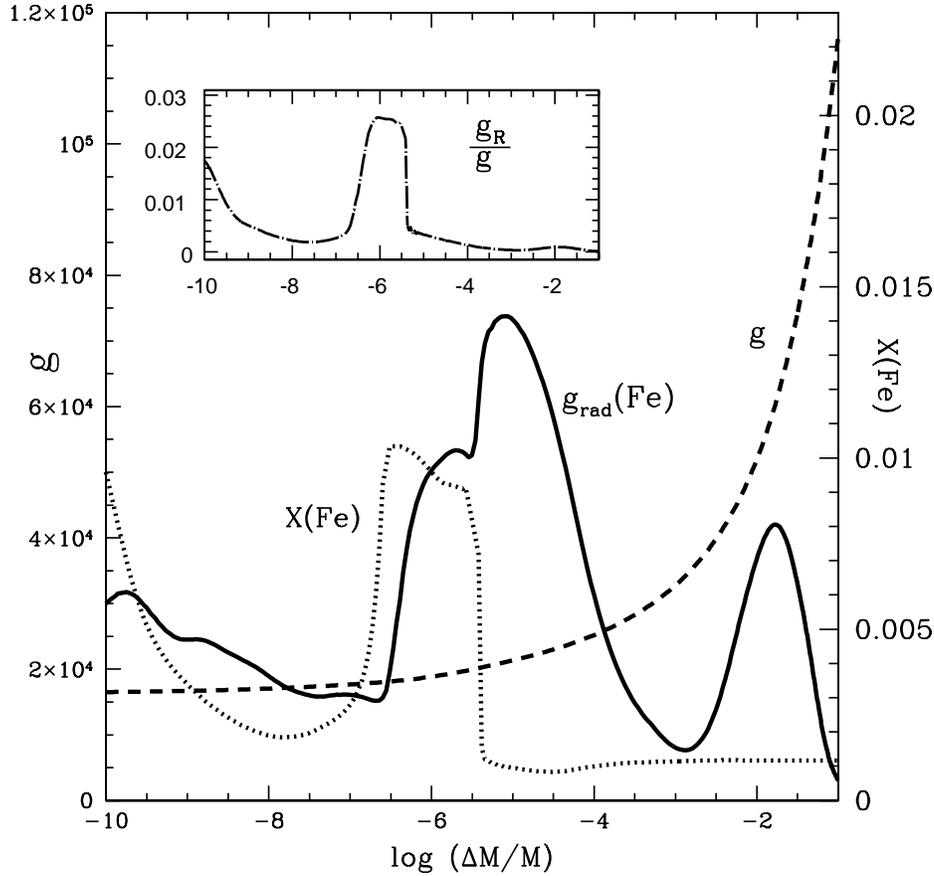}
\caption{Radiative acceleration on iron (large frame) and total radiative acceleration on
the stellar medium (small frame) compared with the local gravity in a 1.7M$_{\odot}$, 403 Myr
model. The dashed curve in the large frame represents the gravity, plotted as a function
of the external mass fraction. The solid curve is the radiative acceleration on iron, which
largely exceeds the gravity in several stellar layers, leading to the iron accumulation. The
dotted curve represents the mass fraction of iron in this model, showing the radiatively induced
accumulation. The shape of this curve is due to the induced iron convective
zone. The small frame presents, with the same abscissa, the ratio of the global radiative
acceleration (g$_R$ = X(Fe)grad(Fe)) induced by iron on the stellar medium, to the local
gravity (after Vauclair and Th\'eado 2012).}
\end{figure}

In particular, the accumulation of iron and nickel, which represent important contributors to the opacity in some stellar layers, leads to a natural local increase of the opacities and consequently to extra convective zones. In some
cases, it may even trigger stellar pulsations through the iron-induced $\kappa$-mechanism (Charpinet et al. \cite{charpinet97}).

\section{Fingering (thermohaline) convection}

Thermohaline convection is a wellknown process in oceanography : warm
salted
layers on the top of cool unsalted ones rapidly diffuse downwards even
in the
presence of stabilizing temperature gradients. When the thermal gradient is large enough to compensate the effect of the $\mu$-gradient, the medium should be stable. However
salted blobs fall down like fingers while unsalted 
matter goes up around. The
reason why the medium is still
unstable is due to the different diffusivities of heat and salt. A warm
salted blob falling down
in cool fresh water sees its temperature decrease before the salt has
time to diffuse
out : the blob goes on falling due to its weight until it mixes with the
surroundings.

\begin{figure}
\includegraphics[width=12.5cm]{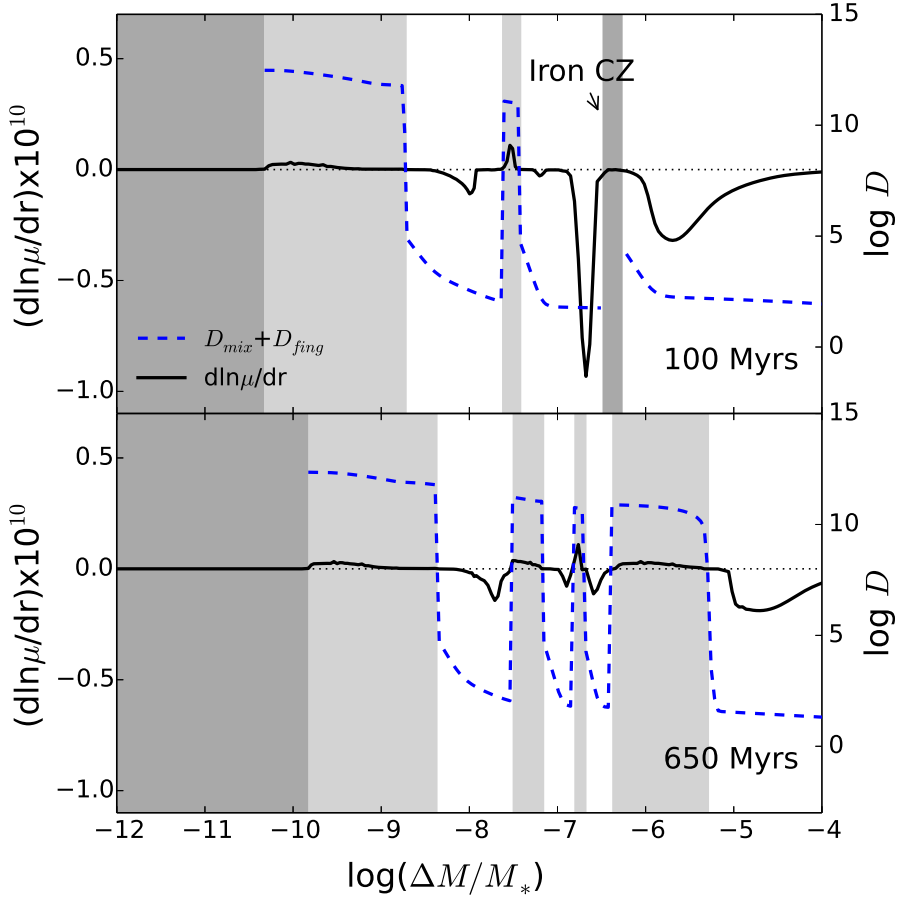}
\caption{Fingering zones which develop in a 1.7M$_{\odot}$ model, presented here at 100 Myr and 650 Myr. The dark grey zones correspond to the dynamical convective regions and the light grey zones correspond to the fingering ones.  The solid lines correspond to the mu-gradients and the dashed lines show the effective mixing coefficients. At 100 Myr the local opacity increase due to the Fe accumulation induces a dynamical convective zone, which is later on replaced by a large fingering convection region (after Deal, Richard \& Vauclair 2016). }
\end{figure}

The condition for the salt fingers to develop is related to the
density variations induced by temperature and salinity perturbations.
Two important characteristic numbers are defined :

$\bullet$ the density anomaly ratio 
\begin{equation}
R_{\rho} = \alpha \nabla T / \beta \nabla S
\end{equation}
 where 
$\alpha = - (\frac{1}{\rho } \frac{\partial \rho}{\partial T})_{S,P}$ 
and $\beta = (\frac{1}{\rho } \frac{\partial \rho}{\partial S})_{T,P}$ 
while
$\nabla T$ and $\nabla S$ are the average temperature and salinity gradients in the
considered zone

$\bullet$ the so-called ``Lewis number" 
\begin{equation}
\tau = \kappa_S/\kappa_T = \tau_T/\tau_S
\end{equation}
where $\kappa_S$ and $\kappa_T$ are the saline and thermal diffusivities
while $\tau_S$ and $\tau_T$ are the saline and thermal diffusion time scales.

The density gradient is unstable and overturns into dynamical convection
for $R_{\rho} < 1$ while the salt fingers grow for $R_{\rho} \geq 1$. On the other hand 
they cannot form
if $R_{\rho}$ is larger than the
ratio of
the thermal to saline diffusivities $ \tau^{-1} $ as in this case the salinity
difference between the blobs
and the surroundings is not large enough to overcome buoyancy.

Salt fingers can grow if the following condition is satisfied : 
\begin{equation}
1 \leq  R_{\rho} \leq \tau^{-1} 
\end{equation}

A similar kind of convection occurs in stellar radiative zones when a
layer with a larger 
mean molecular 
weight sits on top of layers with smaller ones. In this case $\nabla_{\mu}$ = dln$\mu$/dln$P$
plays
the role of the
salinity gradient whereas the difference $\nabla_{ad} - \nabla$
(where 
$\nabla_{ad}$ and $\nabla$ are the usual adiabatic and local (radiative)
gradients 
dln$T$/dln$P$) plays the role
of the temperature gradient. The medium is dynamically unstable if:
\begin{equation}
\nabla_{crit} = \frac{\phi}{\delta}\nabla_{\mu} + \nabla_{ad} - \nabla < 0  
\end{equation}
where $\phi=(\partial$ ln $\rho/\partial$ ln $\mu)$ and $\delta=(\partial$ ln $\rho/\partial$ ln $T)$.
In the reverse case, it is stable against dynamical convection but then thermohaline convection may take place as in the ocean. In stars, this convection is now generally referred to as ``fingering convection".

Several 2D and 3D numerical simulations have been performed, with the aim of giving prescriptions for the computations of fingering convection in 1D stellar models. Let me quote the most recent ones, Brown, Garaud \& Stellmach \cite{brown13}, who gave robust expressions of the mixing coefficients due to fingering convection, and Zemskova et al. \cite{zemskova14}, who precisely simulated the stellar situation, when an element accumulates at some layer due to an external effect (e.g. the radiation flux), leading to fingering instability. They found that a steady state is possible, in which the element accumulation may still be important. This simulation was used later on by Deal, Richard \& Vauclair \cite{deal16} for the case of a 1.7 M$_{\odot}$ star. They showed that fingering convection induced by atomic diffusion leads to large extra-mixing inside such stars (Fig. 2). When this mixing is taken into account, with the hypothesis that the mixing zones are linked together, the observed abundances are well reproduced.

\section{Rotational induced mixing}

Chemical gradients generally develop in stellar interiors, especially for slowly rotating stars. The extra-mixing induced by fingering convection, a process which has long be by-passed by stellar physicists, should always be tested in that respect.

In real stars however, several processes work together, which makes the situation very complex. Rotational induced mixing clearly plays an important role, as the stars which show abundance variations at their surfaces are well known to be slow rotators, whereas the rapid rotating ones do not show such anomalies. Treating the interplay between fingering convection and rotational induced mixing is not trivial. 

Jean-Paul Zahn spent a non-negligible part of his career to work on the problem of rotational induced mixing. During a very productive workshop in Santa Barbara, in 1990, he realized that the mixing induced by meridional circulation could be treated as a diffusive process, provided that the stellar gas be horizontally mixed in an efficient way. This lead to the wellknown paper Zahn \cite{zahn92} and later on Maeder \& Zahn \cite{maeder98}. With several collaborators, including Suzanne Talon, Corinne Charbonnel, Stephane Mathis, he solved the question of the transport of angular momentum associated with the transport of chemical species. They realized however that angular momentum should also be transported by other means to explain observations like the solid internal rotation of the Sun (Talon \& Charbonnel \cite{talon98}, Charbonnel \& Talon \cite{charbonnel99}, Mathis \& Zahn \cite{mathis04}, etc.).

The question of rotational induced mixing in the presence of $\mu$-gradients was first addressed by Mestel \cite{mestel53}, who wanted to explain why the meridional circulation does not penetrate the stellar cores, which would completely change the stellar evolution times. He showed how the presence of vertical stabilizing $\mu$-gradients could be transformed by the circulation into horizontal chemical gradients which, in turn, would induce stellar currents opposite to the circulation. He called ``$\Omega$-currents" the traditional meridional circulation ones, and ``$\mu$-currents" those due to the chemical gradients. He showed how this could oppose the circulation at the limit of the stellar cores.

Later on, the question of the competition between ``$\mu$-currents" and ``$\Omega$-currents" was discussed for the case of the $\mu$-gradients induced by helium gravitational diffusion. It was shown that in this case the ``$\mu$-currents" are generally not negligible and should be taken into account in the computations (Vauclair \cite{vauclair99}, Vauclair \& Th\'eado \cite{vauclair03}, Th\'eado \& Vauclair \cite{theado03a} and \cite{theado03b}, Palacios et al. \cite{palacios03}). Their effects depend however on the chosen value for the horizontal mixing coefficient, which is not really known. 

At the present time, a new situation does appear. The accumulation of heavy elements in some internal stellar layers leads to important inverse $\mu$-gradients, which revives the question of ``$\mu$-currents" in a different way. In this case, the induced horizontal gradients are opposite to the ones discussed in previous works, which means that the ``$\mu$-currents" are now working in the same direction as the ``$\Omega$-currents". Premiminary computations show that it could be non-negligible in the problem of extra-mixing inside stars. This should be studied in a near future.

\section{Conclusion}

Jean-Paul Zahn was a great astrophysicist, who made important steps in undertanding astrophysical fluid dynamics and mixing processes inside stars. He was also quite interested in the questions of atomic diffusion, its consequences and competition with macroscopic processes. The detailed observations of stellar abundances may lead to precise constraints on these processes. Helio and asteroseismology have also totally modified our approach of stellar physics in the sense that it provides deep tests of stellar interiors. 

The mic-mac connexion, which considers the importance of microscopic processes for macroscopic consequences, is important in that respect. Much work still has to be done, which should be undertaken by the new generations of astrophysicists.


\end{document}